\documentclass[12pt]{article}
\usepackage{epsfig}
\begin{document}

\begin {center}
{\bf  {GALAXIES AS CONDENSATES}}
\vskip 5mm
{\rm D.V. BUGG \footnote {email: david.bugg@stfc.ac.uk}} \\
{ Department of Physics, Queen Mary, University of London, London E1\,4NS,
UK} \\
\vskip -1mm
\end {center}

\begin{abstract}
\noindent
A novel interpretation of MOND is presented. 
For galactic data, in addition to Newtonian acceleration, there is an 
attractive acceleration peaking at Milgrom's parameter $a_0$.
The peak lies within experimental error where $a_0 = cH_0/2\pi$ and
$H_0$ is the present-time value of the Hubble constant.
This peaking may be understood in terms of quantum mechanical mixing
between Newtonian gravitation and the condensation mechanism.  
There are five pointers towards galaxies being  Fermi-Dirac condensates.  

 %{Keywords: Galaxies; MOND }
\end{abstract}

\section {Introduction}
The objective of this paper is to introduce a model which gives a
physical interpretation of MOND.
The most relevant astrophysical data will be discussed.
They contain five clues which suggest that galaxies
are quantum mechanical condensates.
These clues fit neatly together.

It is well known that the rotation curves of galaxies deviate from
Newtonian mechanics.
The question is why?
Famaey and McGaugh have recently provided a review of all aspects
of the data, with an exhaustive list of references \cite {Famaey}.
Milgrom's approach should need little introduction. 
He proposed in 1983 a modification of Newtonian mechanics \cite {MilgromA},
\cite {MilgromB},
where the observed total acceleration $a$ depends on the Newtonian 
acceleration $ g_N$ as
%Eq. (1)
\begin {equation}
a = g_N/\mu (\chi), 
\end {equation}
and $\chi = a/a_0$; $a_0$ is a constant.
For small $g_N$, all forms in use for $\mu$ go at large radius $r$ to
%Eq. 2 
\begin {equation}
a = \sqrt{ g_Na_0}.
\end {equation}
A star with rotational velocity $v$ in equilibrium with centrifugal 
force satisfies
%Eq. (3)
\begin {equation}
v^2/r = \sqrt{ \frac {GM}{r^2}a_0}; 
\end {equation}
here $G$ is the gravitational constant and $M$ the mass of the 
galaxy; $r$ cancels and 
%Eq. (4) 
\begin {equation}
v^4 = a_0 GM.
\end {equation}
It is striking  that the $v^4$ relation to $M$ agrees well with 
the empirical Tully-Fisher relation between observed velocities at the 
edges of galaxies and the luminosity \cite {Tully}.
McGaugh shows that this relation applies over 6 decades of galactic
masses from $\sim 10^6$ to $10^{12} M_\odot$ after including the mass of
gas and dust in each galaxy \cite {McGaughA}.
It is important that in MOND there is only one free parameter $a_0$, 
fitted to all galaxies once a particular form for $\mu(x)$ has been chosen.
This is the order parameter of the condensate.

Within the last few years, an independent observation of the same
phenomenon has appeared in globular clusters.
These spherical clusters of stars have dimensions of a few parsec
(pc), i.e. a factor $\sim 10^4$ smaller than the Milky Way.
Scarpa et al. reported initially on two globular clusters situated 16--19 kpc 
from the Milky Way \cite {Scarpa}.
The equilibrium of such clusters is controlled by Jeans' Law, which
relates the velocity dispersion of stars to their acceleration.
Scarpa et al. traced the velocity dispersion of 184 stars at large
radius, identified as being members of one globular cluster (rather than 
interlopers), and 146 stars in the second cluster.
The velocity dispersion is maximal at the centre of each cluster.
They were able to trace it to a radius $r_0$ 
twice that where Newtonian acceleration reaches $a_0$.
Velocity dispersions deviate rather abruptly from Newtonian 
acceleration as it decreases through $a_0$.
Tidal heating by the Milky Way varies as $r^{-3}$, and is at least one 
order of magnitude smaller, making its effect negligible.

Scarpa et al. have made observations of a further 6 globular clusters.
Hernandez and Jim\' enez give the algebra relating velocity dispersions
of stars to Newtonian acceleration using Jeans' Law 
\cite {HernandezA}.
Hernandez, Jim\' enez and Allen report a detailed study of the velocity
dispersion profiles of all 8 globular clusters \cite {HernandezB}.
Like Scarpa et al., they conclude that tidal effects are significant only
at radii larger by factors 2--10 than the radius where MOND flattens
the curves.
They also show that the velocity dispersion $\sigma$ varies with the mass
$M$ of the cluster as $M^{-4}$ within errors; this is the expected
analogue of the Tully-Fisher relation arising from Jeans' Law.

This result is independent of luminosity measurements used in 
interpreting galactic rotation curves.
In galaxies, the mass $M$ within a particular radius is not easy to
determine, and is usually taken as the mass where rotation curves
flatten out.
Further exploration of globular clusters is highly desirable. 
It has not been suggested till now that Cold Dark Matter has
any significant connection with globular clusters.

Several forms for the function $\mu$ are used today.
Effectively they are mappings and the question is how to interpret them.
Differences between them are illustrated in Fig. 19 of the review of
Famaey and McGaugh \cite {Famaey}.
A particular form due to Milgrom \cite {MilgromC}
%Eq. (5) 
\begin {equation}
\mu (\chi) = \sqrt {1 + 1/(4\chi^2)} - 1/(2\chi)
\end {equation}
will be used here.
It has the smoothest variation with $\chi$ and is also symmetric about $a_0$.
At present only 8 globular clusters have been considered, but the
algebraic form with which they are fitted is closely consistent with this
equation. 
Within individual errors ($\sim 40\%$), results confirm 
Eq. (5) for accelerations above $a_0$, where galaxies are subject to 
systematic error in the variation of integrated mass as a 
function of radius. 

The standard cosmological model $\Lambda CDM$ assumes that condensation
of galaxies is a two-stage process. 
Dark Matter, whose nature is not yet identified, condenses gravitationally
and provides a host in which galaxies condense as a second step.
The view adopted here is that quantum mechanics replaces Dark Matter
via a one-step process. 
This should not be surprising. 
The gravitational field is carried by gravitons and quantum mechanics
plays a crucial role in forming Black Holes.

The parameter $a_0$ of Milgrom's model agrees within experimental error
with $cH_0/2\pi$, where $H_0$ is the present-time value of the Hubble
constant.
However, the connection is subtle and appears only at the final step.
A further clue is that the phenomenology of galactic rotation curves
appears on a log-log plot. 
This will be associated here with the Partition function of Statistical
Mechanics.
Later, it will be associated with the idea that quantised bosons or fermions
play a role in forming a quantum mechanical condensate that {\it requires}
that observed features appear on a log-log plot.
The Partition functions for quantised bosons or fermions are
%Eq. (6) 
\begin {equation}
Z = \Pi _s \,\frac {1}{1 \pm e^{E_s/kT}},
\end {equation}
where $E_s$ are energies of levels in a box and $\Pi$ denotes the product of
the numbers of all possible quantum levels in the box.
The $+$ sign gives Fermi-Dirac statistics and the $-$ sign Bose-Einstein.
For a refresher course, see Schr\" odinger's very clear exposition
\cite {Schrodinger};
$\log Z \to \sum_s E_s/kT$ in the classical limit, $k$ is Bolzmann's 
constant and $T$ the temperature.

%Fig. 1
\begin{figure}[htb]
\begin{center} \vskip -15mm
\epsfig{file=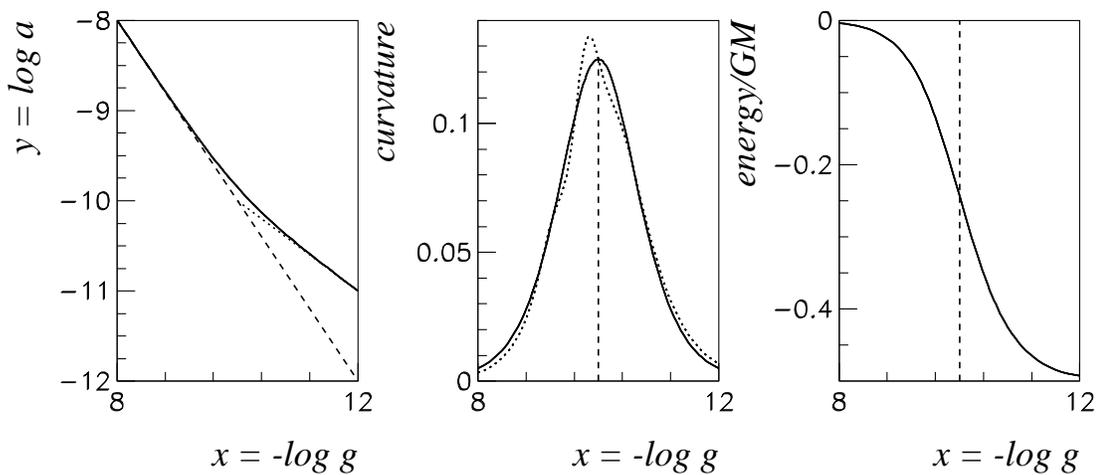,width=16cm}
\vskip -6mm
\caption
{(a) The full curve shows the result of equation (10); the dashed  
line Newtonian acceleration,
and the dotted one a straight line given by the term $\sqrt {a_0g}$;
(b) full curve: the peak arising from curvature of the full curve in (a); 
the dotted curve shows a fit discussed in Section 4; 
(c) full curve: the energy  derived from the full curve of (b).}
\end{center}
\end{figure}
Milgrom plots the log of total acceleration against the log of 
gravitational acceleration for systems of different masses \cite {Milgrom9}. 
This presentation is followed in figure 1(a).
Axes are $x={\rm log}_{10}\, g_N$, where $g_N$ is Newtonian acceleration,
and $y={\rm log}_{10}\, a$, where $a$ is total acceleration.
Newtonian acceleration follows the diagonal dashed line.
The full acceleration has a bend centred at $a_0$.
The value of $a_0$ is close to $10^{-10}$ m\, s$^{-2}$ 
and it simplifies figures and arithmetic to adopt this value.
The dotted line of Fig. 1(a) follows the asymptotic form $\sqrt {a_og}$ 
of the acceleration.
In earlier work, attention has been mostly to the asymptotic form of the
dotted line on Fig. 1(a).
Astrophysicists have used this to determine the parameter $a_0$.
The dashed and dotted curves meet at acceleration $a_0$.
This suggests some form of cross-over between Newtonian acceleration
and the acceleration $\sqrt {a_0g}$.

The scheme introduced here is to focus attention on the {\it extra}
acceleration which is additional to Newtonian acceleration on Fig. 1(a),
i.e. the difference between the full curve and the 
dashed line for Newtonian acceleration.
A detail is that effects of General Relativity in a galaxy the size of
the Milky Way are at a level of $(2 \times 10^{-4})g_N$;
this is well below experimental errors. 
This tiny correction is neglected here and simplifies formulae greatly,
[Later, it could be re-introduced as a perturbation.]
The additional acceleration is taken as $d^2y/dx^2$ of Fig. 1(a) and is
shown enlarged on Fig. 1(b) as the full curve.  

Section 2 works through the algebra of Figs. 1(a) and (b) and shows
that galaxies have considerable stability.
Subsection 2(a) examines other forms of Milgrom's function $\mu (x)$,
with the objective of assessing errors in the acceleration fitted to
Fig. 1(b).

Section 3 discusses the total extra energy $W$ obtained by integrating the
non-Newtonian component over x.
The way the algebra is done takes into account explicitly the exponential
in the Partition function. therefore parametrising the total acceleration
exactly.
However, it is necessary to realise that results appear on a logarithmic
horizontal scale.
Newtonian gravity is normally expressed as a function of $r$;
in equations which appear later, it will be transformed to a function of $x$.
This is why Milgrom's function $\mu$ has been described as a mapping.
It is a matter of convention that the zero of Newtonian energy is taken
at infinity (but ignoring the effect of Dark Energy over the Universe
as a whole).
Gravitational energy $E$ is then negative for finite radii $r$.
The extra energy associated with non-Newtonian gravity has the same
negative sign as Newtonian energy.
It is shown on Fig. 1(c).
It turns out that the extra energy at infinity is $-0.5 GM$, where $G$
is the gravitational constant and $M$ is taken as the mass of the galaxy
integrated from zero radius to the radius where the total acceleration is
$a_0$.
This will be interpreted as an energy gap arising from the formation of
a Fermi-Dirac condensate.
The factor of 0.5 arises directly from the factor 2 difference
in slope of the dotted and dashed lines on Fig. 1(a).

In order to explain the origin of the energy gap, it is desirable to
make contact with techniques used in quantum mechanics to understand energy 
levels in nuclei and mesons.
These were developed independently by Bogoliubov \cite {Bog} and
Valatin \cite {Valatin}.
They introduced the idea of working in axes which rotate Fig. 1(a)
anti-clockwise by the mean angle of the dashed and dotted lines with respect
to the $x$ axis; the curve is then symmetrical like the upper curve of Fig. 3.
The solution for total energy is given by the well known
Breit-Rabi equation \cite {Breit}.
There are reasons given later why the lower curve of Fig. 3 is unlikely to
appear in galactic rotation curves.

The quantum mechanical mixing between gravity and the Hubble acceleration
is discussed in Section 3.1.
A consequence of the asymptotic form $\sqrt {a_0g_N}$ fitted to galactic
rotation curves is that it produces a weak tail to the Newtonian
potential with a logarithmic dependence on radius $r$. 

Condensates are well known in Particle Physics and are reviewed
briefly in Section 4. 
One of them, Chiral Symmetry Breaking, appears to proceed in close
analogy with the Fermi function describing galaxies.
The interpretation given to Fig. 1(c) is that the graviton develops
an effective mass over the curved section of Fig. 1(a).
Its parameters are derived with errors of $\pm 8\%$ of $GM$.
Exactly how this effective mass arises is beyond the scope of the
present paper and requires the technical expertise of theorists
familiar with renormalisation effects and gauge invariance, but a suggestion 
is made how this may go.
Section 5 reviews further work needed in interpreting galaxies and 
Section 6 draws conclusions.

\section {\small {THE MODEL}}
From equation (5), 
%Eqs. (7-9) 
\begin {eqnarray} 
\mu = g_N/a &=& \sqrt {1 + \frac {a^2_0}{4a^2}} - \frac {a_0}{2a} \\
( g_N/a + a_0/2a) ^2 &=& 1 + (a_0/2a)^2 \\
g_N^2 + a_0g_N &=& a^2.
\end {eqnarray}
The cancellation of the two terms $(a_0/2a)^2$ is typical of a dispersion
relation where a subtraction may be made at any value of $a$.
This emerges as an important point in Section 4.

It is inconvenient to use axes to the base 10, so they will be
replaced in the algebra which follows by conversion to $\rm {ln}_e$. 
Where necessary in comparing with Fig. 1, the conversion factor
2.3026 will be restored. From Eq. (9) 
%Eqs. (10-11)
\begin {eqnarray}
y  & = & \ln _e \sqrt {g^2 + a_0g}; \\
x  & = & \ln _e \, g,
\end {eqnarray}
where the suffix is dropped from $g_N$.
From equation (10),
%Eqs. (12-15)
                                                                                                                                                                                                                              \begin {eqnarray}
e^{2y} &=& e^{2x} + a_0 e^x \\
dy/dx &=& (e^x + a_0/2)(e^x + a_0)^{-1} \\
d^2y/dx^2 &=& (a_0/2)e^x(e^x + a_0)^{-2} \\
d^3y/dx^3 &=& (a_0/2)e^x(a_0 - e^x)/(e^x + a_0)^{-3};
\end {eqnarray}
$d^3y/dx^3$ goes to zero at $g = a_0$.
[There are higher order terms too.]
The curvature $d^2y/dx^2$ has a maximum value at $x = a_0$, where
$(d^2y/dx^2) = 1/8$.
The starting equation (5) gives the smoothest symmetric
curve for $d^2y/dx^2$ from the alternative forms for Milgrom's 
$\mu (\chi )$.

In real galaxies, perturbations arise from thermal and pressure 
effects. 
Some narrow structures are observed in rotation curves, for example from 
bars in large galaxies. 
However, these may be fitted empirically using Poisson's equation for 
mass structures appearing in the gravitational potential.
The MOND component `rides' such structures smoothly, see Figs. 21 and 29
of Famaey and McGaugh \cite {Famaey}.
An immediate question is why Planck's constant does not appear in 
results for galactic rotation curves. 
The reason is that galaxies are noisy enough to hide it.

Using the known value of $H_0$, the maximum curvature is expected at 
$a_0 = (1.113 \pm 0.046) \times 10^{-10}$ m\, $s^{-2}$, 
i.e. $\log _{10} a_0 = -9.953$.
Results are insensitive to the small difference from $10^{-10}$.
McGaugh summarises a large number of papers comparing the Tully-Fisher 
relation with models of galaxy formation \cite {McGaugh11}.
He concludes that gas rich galaxies give the best determination of
the baryonic masses of galaxies: $a_0 = (1.3 \pm 0.3) 
\times 10^{-10}$ m\, $s^{-2}$.
A slightly lower value $1.22 \pm 0.33 \times 10^{-10} \, {\rm m}\, 
s^{-2}$ is found by Gentile, Famaey and de Blok \cite {Gentile}. 
Fig. 1(b) shows $d^2y/dx^2$ as a peak $dW/dx$ in the acceleration at $a_0$,
marked by the dashed line;
here $W$ is the energy of the `extra' contribution.
The full curve is well approximated by a Gaussian for the acceleration:
%Eq. (16) 
\begin {equation}
d^2y/dx^2 = 0.125\exp -[\gamma (x - a_0)^2]
\end {equation}
with $\gamma = 1.175$,
i.e. the Gaussian drops to half-height at $9.6\%$ of the value of $x$
at the dip. 
The conclusion is that galaxies have considerable stability.

\subsection {Alternative forms for $\mu$}
Other forms for $\mu (\chi)$ have been used to fit galactic rotation 
curves. Their effects are rather small.
Two common examples are (A) $\mu (\chi) = \chi /(1 + \chi)$, 
(B) $\chi /\sqrt {1 + \chi ^2}$.
Both forms have the effect of moving the peak of the curvature slightly: 
from $x = -10$ to $-9.7$ for A and to $-9.85$ for B.
The height of the peak for A is scaled by a factor 0.78 and the curve
becomes correspondingly wider, resulting in a tail reaching 0.012 at
$x = -8.2$ and $-11.8$. 
The result for Fig. 1(c) is that the top of the Fermi function is 
$4\%$ of $GM$ lower, and the bottom of the curve higher by the same amount, 
but the central part of the Fermi function is unchanged and still centred very
close to  $x = -10$.
Form B produces the converse effect: a higher, narrower peak in Fig. 1(b)
and a Fermi function beginning closer to the top and finishing closer to
the bottom of Fig. 1(c), but with its central section undisturbed.

A detail is that one might consider the possibility that the smooth curve
of Fig. 1(a) could be replaced by the dashed and dotted lines.
This would give rise to a sharp cusp in Fig. 1(b).
Such cusps are known in Particle Physics, but arise only at the
opening of phase space for a new reaction channel \cite {bugg}.
However, no such threshold exists in galactic phenomena.

The Gaussian acceleration of Fig. 1(b) can be integrated to determine the 
corresponding energy using the Error function or the closely related Normal 
Probability Integral as described in items 590 and 585 of Tables of 
Integrals of Dwight \cite {Dwight}.
The result is shown by the full curve in Fig. 1(c).

%Fig. 2
\begin{figure}[htb]
\begin{center} \vskip -16mm
\epsfig{file=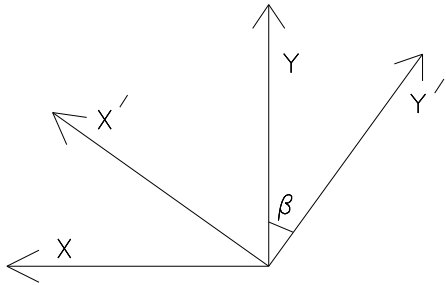,width=11cm}
\vskip -45mm
\caption
{Axes $x$, $y$,\, $x'$ and $y'$.}
\end{center}
\end{figure}
At this point a rotation of axes is introduced with the objective of making
the curve of Fig. 1(a) resemble Fig. 3. 
This is the Bogoliubov-Valatin transformation mentioned earlier.
It has the effect of rotating the full curve of Fig. 1(a) through the
mean angle of the dashed and dotted lines.
This is a standard transformation where there is quantum mechanical
mixing between any two basis states.
The rotation is about the point $x = -10$, $y = -10$ where the two
straight lines of Fig. 1(a) cross:
%Eqs. (17-18)
\begin {eqnarray}
x' &=& (x+10)\cos \beta - (y + 10) \sin \beta \\
y' &=& (x+10)\sin \beta + (y+10)\cos \beta.
\end {eqnarray}
Substituting the results of Eqs. (10)--(15) gives an exact
expression for the curve in $x',y'$ axes.
It is also convenient to re-express Newtonian acceleration and energy directly 
in terms of $x$, with the following simple manipulations; from Fig. 1(a) 
%Eqs. (19-21)
\begin {eqnarray}
g &=& e^x = GM/r^2 \\
e^{x/2} &=& \sqrt {GM}/r \\
\frac {dr}{r} &=& - \frac {dx}{2}. 
\end {eqnarray}
The Newtonian potential is
%Eq. (22)
\begin {equation}
E_1 = -GM/r =-\sqrt {GM}e^{x/2}.
\end {equation}

\section {\bf The relation to a Fermi function}
The acceleration differs in $x$ and $x'$ axes, but the scalar quantity
$W$ is independent of the choice of axes; this is used as a check on
algebraic manipulations. 
The simplest form for the Fermi function is to take
\begin {equation}  
%Eq (23)
W(x') \propto \left[ 1 + \exp \left(
\frac {E - E_F}{kT}\right) \right]^{-1},
\end {equation}
where $E_F$ is the energy at the centre of the Fermi function and $T$ is an
adjustable unknown temperature.
This gives the full curve shown in Fig. 1(c).

It is also of interest to explore how stable this solution is.
This has been studied in a number of different ways; the one giving the
clearest results will be presented here.
For this purpose, it is useful to express the energy distribution including a
parabolic dip near temperature $T$, hence allowing for the possibility 
that a reduced temperature is appropriate there, corresponding to the
existence of a condensate. 
Minor details are needed 
(a) to introduce a small offset $x_0$ arising from the term $a_0\sqrt {g_N}$, 
(b) an exponential form factor to cut off the parabolic dip in the denominator. 
The final form is 
%Eq. (24) 
\begin {equation}
W(x') = -0.5/\left[ 1.0 + \exp \left( \frac {1-\gamma '(x'+ x_0)}
{1 + \alpha (x'+ x_0)^2\exp [-\eta (x' + x_0)^2]}  \right) \right];
\end {equation}
$W(x')$  simply adds to the Newtonian potential. 
The energy range of the Fermi function in Fig. 1(c) is
%Eq. (25) 
\begin {equation}
|\Delta E| = 0.5\, GM.
\end {equation}
This is a large effect locally.
As mentioned in the Introduction, it is not accidental that the numerical 
result comes out to be $0.5\, GM$. 
This can be traced directly to the asymptotic form $\sqrt {a_0g}$ of MOND, 
i.e. slope 0.5 on Figure 1(a).

%Fig. 3
\begin{figure}[htb]
\begin{center} \vskip -16mm
\epsfig{file=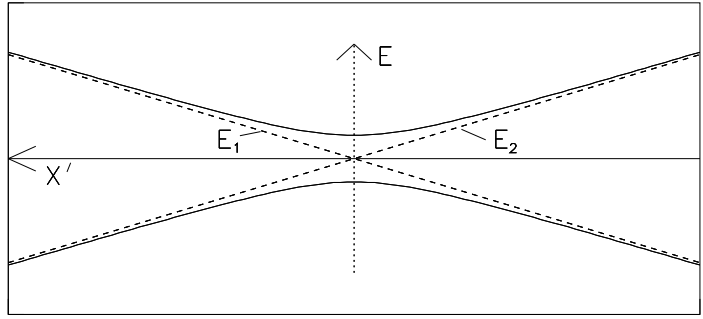,width=13cm}
\vskip -10mm
\caption
{Sketch of two crossing atomic lines; full lines show them including
mixing, dashed lines without; $E_1$ and $E_2$ label the convention for 
eigenvalues in the case of no mixing.}
\end{center}
\end{figure}

For small $\alpha$, i.e. a small parabolic dip, the asymmetry of the
fit is very small, as displayed by the full curves of Figs. 1(a) and
(b). 
At the other extreme, the dashed curve of Fig. 1(b)  illustrates the effect
of $\alpha = 0.6$, which increases the 
$\chi^2$ of the fit to Fig. 1(b) by a factor 10; acceleration is less well
determined than the Fermi function because it is a gradient term. 
The effect on the fit to Fig. 1(c) is no larger than the line width.
The largest changes observed in all tests are $\pm 4\%$ of the height
of the Fermi function at both top and bottom.
One should remember that for real galaxies, asymmetries of this variety
can be due to the decrease in the relevant mass $M$ inside radius $r$.

\subsection {\it Quantum mechanical mixing between gravity and the condensation
mechanism.}
The Breit-Rabi equation was first derived to describe crossing atomic 
lines in a magnetic field \cite {Breit}. 
Where they cross, there is quantum mechanical mixing between them if basis states
are not orthogonal.
Fig. 3 sketches curves and 
eigenvalues $E_1$ and $E_2$ for atomic levels as a function of magnetic 
field.
Away from the crossing point, the eigenvalues vary linearly with 
magnetic field, but repel one another where they cross unless the mixing
potential $V = 0$.
The interpretation put on Milgrom's formula is that there is
distinctive mixing. 
The maximum repulsion between levels is at the crossing point.
As pointed out in the Introduction, 
Fig. 1(a) is a similar diagram turned $35.8 ^\circ$ clockwise. 

Because galaxies are large ensembles of stars, gas and dust, it is necessary 
to take expectation values of quantum mechanical operators.
The place where Quantum Mechanics is needed is to 
describe the mixing between the condensation energy and Newtonian
energy and the related energy eigenvalues. 
The eigenvalue equation defines the mixing potential $V$ fitted to data:
%Eq.  (26)
\begin {equation}
<H \Psi>  = \left(
\begin {array} {cc}
<H_{11}> & V \\
V & <H_{22}>
\end {array} 
\right)
\left( \begin {array} {c}
\Psi _1 \\
\Psi _2
\end {array} 
\right)
= E\left( \begin {array} {c}
\Psi _1 \\ 
\Psi _2 
\end {array}
\right) .
\end {equation}
%Eqs. (27)-(29)
\begin {eqnarray}
H_{11}\Psi_1 + V\Psi_2 &=& E\Psi_1 \\
H_{22}\Psi_2 + V\Psi_1 &=& E\Psi_2 \\
(H_{11}-E)(H_{22}-E)-V^2 &=&0.
\end{eqnarray}
We work in the galactic centre of mass and in axes rotating
with it.
The rotational energy is ignored; if it is included,
it simply describes the rotational equilibrium.
$H_{11}$ and $H_{22}$ are mean values summed over the ensemble.

Data on galactic rotation curves require that they follow the same
behaviour over several orders of magnitude in mass from large ellipticals
to dwarf galaxies. 
Then 
%Eqs. (30)-(32)
\begin {eqnarray}
H_{11} &=& E_1 = -GM/r = -\sqrt {GM}e^{x/2}\\
H_{22} &=& E_2 = \sqrt {GM}\epsilon (x) \\
V      &=& \sqrt {GM}W(x'),
\end {eqnarray}
where $\epsilon$ refers to an energy possibly given by the Hubble
acceleration. 
The factor $\sqrt {GM}$ appears in all three equations.
The common approximation that the gravitational field of a disc galaxy is 
$GM/r$ is also used, where $M$ is the mass inside radius $r$.
This is accurate to $\le 1\%$ at the large values of r of interest.
If the variation with $r$ is known experimentally, detailed information on 
the distribution could easily be adopted for each individual galaxy.

The two solutions for the Breit-Rabi equation are   
%Eq. (33)
\begin {equation}
E = \frac {E_1 + E_2}{2} \pm \sqrt{\left( \frac {E_1 - E_2}{2} \right) ^2 +
V^2}.
\end {equation} 

\subsection {A long-range logarithmic tail to the Newtonian potential}
If there were no quantum mechanical mixing between Newtonian acceleration
and the condensation mechanism, there would be no structure at the crossing
point and, more serious, no explanation for the term $\sqrt {a_0g_N}$. 
 
Asymptotically, the total acceleration, taken from MOND, is
%Eq. (34)
\begin {equation}
a = a_0 \sqrt {g_N/a_0}.
\end {equation}
Since $g_N = GM/r^2$,
%Eq. (35) 
\begin {equation}
a \to \sqrt {GMa_0}/r.
\end {equation}
Taking this as $-d\phi /dr$, where $\phi$ is a potential induced
by the mixing,
%Eq. (36) 
\begin {equation}
\phi \to -\sqrt {GMa_0}\ln (r/r_1).
\end {equation}
Here $r_1$ is the mean radius for this
term. 
The value of $r_1$ is necessarily very close to the much larger
dip caused by $dW/dx'$.
Because $a_0 \sim 10^{-10}$, this term is very small.
However, it does explain the asymptotic straight-line at the right-hand
edge of Fig. 1(a).

The interpretation of this term is simple.
Mixing between the Newtonian potential and the condensation mechanism
allows the wave-length of gravitons trapped in the Newtonian
potential to expand. 
This lowers the zero-point energy.
An analogy is the covalent bond in chemistry.
In the hydrogen molecule, each of two electrons is attracted
to two protons (which themselves repel one another).
This increases the wave-length of the electrons and
reduces their zero-point energy. 

Sobouti noticed this long-range logarithmic
component and wrote two closely related papers \cite {SiboutiA},
\cite {SiboutiB}. 
These included effects of General Relativity.
This is interesting but not strictly necessary at present.
The problem was solved by Sobouti as a power series with additional
empirical terms proportional to $1/r$ and $1/r^2$.
These terms are replaced by our equations.

There is evidence that the Milky Way has a halo extending to O(100 kpc)
\cite {Deason}  and \cite {Gupta}.
However, this could be due to gas and dust shared with the local cluster
of galaxies.
The halo can also be explained by Cold Dark Matter.
So this halo is presently ambiguous. 

\subsection {Solving the Breit-Rabi equation}
The binding energy $W$ is defined with the same sign as 
$E_1$. 
From equation (31), 
%Eq. (37-39) 
\begin {eqnarray}
2E &=& E_1 + \epsilon (x) \pm 
\sqrt {(E_1-\epsilon (x))^2  + 4W^2} \\
&=& \frac 
{\sqrt {GM}}{\rm {ln} _e {10}}
\left( -e^{x/2}+\epsilon (x) 
\pm \sqrt {(e^{x/2}-\epsilon (x))^2 + 4W^2}\right) \\
2\frac {dE}{dx} &=& \frac 
{\sqrt {GM}}{\rm {ln} _e {10}}
\left( -0.5e^{x/2} +d\epsilon /dx \pm 
\frac {(e^{x/2}-\epsilon (x) )(0.5e^{x/2}-d\epsilon (x)/dx) + 4 WdW/dx}
{\sqrt {(e^{x/2} - \epsilon (x))^2 + 4W^2}} \right)
\end {eqnarray}
The factor $\rm {ln} _e {10}$ is included to conform with the fact that
Fig. 1(a) has been drawn using axes which use logarithms to the base 10.
For the upper branch of the solution, the minus sign for the term involving
the square root is required to reproduce the usual Newtonian potential.

%Table 1
\begin{table}[htb]
\begin {center}
\begin{tabular}{ccc}
\hline
Parameter & fit 1 & fit 2 \\\hline
$\beta$ & $35.78^\circ$ & $35.78^\circ$\\
$\gamma' $  & 1.852     & 1.864 \\
$\eta $ &  $\ge 4.6$    & $\ge 4.6$\\
$x_0$ & 0.600 & 0.572 \\
$\alpha$ & 0.0 & 0.8  \\
$\epsilon$ & 0 & 0  \\\hline
\end{tabular}
\caption{Fitted parameters. }
\end{center}
\end{table}
In practice, the Hubble expansion varies over the radius of the Milky Way
only by a factor $\sim 2 \times 10^{-4}$.
As a result, $\epsilon$ optimises at zero.
Fitted parameters are shown in Table 1 for two
cases.
The best fit is with $\alpha \simeq 0$, i.e. negligible curvature of the dip,
and the second gives a $\chi^2 $ worse by a factor 10 for purposes of
illustration. 
The first gives the full curves of Figs. 1(b) and (c) and the second gives
the dashed curve of Fig. 1(b).
The parameter $\gamma '$ is derived from $\gamma$ of Fig. 1(b), but is
larger because of the rotation of axes from $x,y$ to $x',y'$,
The value of $\eta$ is expected to be large, so that the curve for
$W^2$ makes a rapid transition to small values near $x = -12$,
where the asymptotic form of $\sqrt {a_0g_N}$ generates the very weak
logarithmic tail of the Newtonian potential.
Numerically, $\eta \ge 4.6$ and its precise value has little effect
on the remainder of the fit.
The value of $\alpha$ is the least well determined parameter; it is 
what determines $T_r$.

A Bose-Einstein condensate does not fit the data. 
In this case, $W$ should vary as $T^{3/2}$ at the bottom of the dip
\cite {Tilley2}. 
For positive $x'$ near $x' = 0$ 
%Eqs. (40-41)
\begin {eqnarray}
W(x') &=& -B(1 + |x'|\,^{3/2}\exp {-\gamma '\, x'{^2}}) \\
dW(x')/dx' &=& B(1.5|x'|\,^{0.5} - 2\gamma '\,|x'|\,^{5/2})
\exp {-\gamma '\, x'\,^2}; 
\end {eqnarray}
for negative $x'$, the opposite sign of $|x'|$ is needed in $W(x')$.
The second term in $dW(x')/dx'$ wrecks the $x'$ dependence, which fails
to fit the observed peak.
Near $x'=0$ it has a square root variation with $x'$ and then, when
the second term of Eq. (40) overtakes it, the curve turns downwards. 
This rules out a Bose-Einstein condensate.

\section {An intriguing connection with Particle Physics} 
It is useful to review briefly condensates which are known 
in Particle Physics, since they offer analogies which may carry over to
galaxies.
The idea of a condensate was introduced by Gell-Mann and  Levy 
\cite {Gell-Mann}.  
They were trying to understand the connection between Weak Interations
responsible for $\beta$ decay and the Strong Interactions.
Parity conservation is violated in Weak Interactions but not in Strong
Interactions.
Feynman and Gell-Mann provided an explanation of the Weak 
Interactions, using an analogy between Electromagnetism and the
Weak interactions \cite {Feynman}.
Today that idea has been expanded to cover all Weak Interactions
and is part of the Standard Model of Particle Physics, see Section 10
of the Particle Data Book \cite {PDG}.
The Weak Interactions contain a Vector current which is exactly
conserved and an Axial current which is nearly conserved but not exactly.
Goldberger and Trieman derived a relation describing the axial current 
based on perturbation theory \cite {Goldberger}, but Gell-Mann and Levy 
were concerned that their derivation suffers from renormalisation problems.
That same problem remains a central issue in gravitation.
They introduced a Lagrangian where both $\pi$ and $\sigma$ fields 
appear in the Lagrangian but with different masses;
the $\sigma$ decays to $\pi \pi$ and is responsible for most of nuclear
binding energy.
The Lagrangian can be transformed so that the $\sigma$ term replaces
the nucleon mass.
Along these lines, Nambu and Jona-Lasinio  introduced the idea that the 
nucleon mass arises from an energy gap like that in the theory of 
superconductivity \cite {Nambu}.
The idea of nuclei as liquid drops is the classic example of a
condensate.

The idea of Chiral Symmetry Breaking was introduced by Gasser and 
Leutwyler \cite {Gasser} and is today used to account for a range
of quantitatively accurate results in meson spectroscopy. 
The first step in understanding the precise mechanism was made by 
Bicudo and Ribiero \cite {Bicudo}.
Their work finds the Bogoliubov-Valatin transformation to be an
essential element.
Since then, the situation has clarified further via Lattice Gauge
calculations. 
The Strong Interaction is mediated by gluons obeying Chiral Symmetry.
Below a mass of $\sim 1 $ GeV, the gluon acquires an
effective mass from its interaction with light quarks, which themselves
have masses of $\sim 4$ and 9 MeV.
Above 1 GeV, there is a cross-over in which Chiral Symmetry is largely
restored and the quark model reigns supreme, though with small amounts
of mixing with meson-meson and/or $qq\bar q\bar q$ basis states.
A recent paper of Pennington and Wilson gives details including figures 
showing the cross-over at 1 GeV \cite {Pennington}.

An intriguing conjecture is that the graviton-nucleon interaction 
follows the same pattern as Chiral Symmetry Breaking and 
develops an effective graviton mass over a narrow range of accelerations 
centred on  $a_0$ and disappearing a little beyond $x = -8$, where the
curvature of Fig. 1(a) becomes negligible.
For $x < \rm {log} \, {a_0}$, there is a smooth transition to the logarithmic 
tail of the potential.

Let us now return to Figs. 1(b) and (c). 
What emerges from the range of fits described by Table 1 is that 
Newtonian acceleration is very small near $x = a_0$
compared with that originating from the extra acceleration from
$dW/dx'$.
This second term dominates by a huge factor $2.1 \times 10^4$ at the 
centre of the curve, where $x = {\rm log}_{10}\, a_0$.
At $x = -8$, this factor falls to 70 and is still falling fast.
The conclusion is that the curved part of Fig. 1(a) is the dominant 
feature and Newtonian gravitation in this region is only a small 
perturbation.

Consider the effect of this result near the centre of the Fermi function 
at $E = E_F$ in Fig. 1(c). 
If we retain only the dominant terms in $W$ and $dW/dx$, the results are
%Eqs. 42-43. 
\begin {eqnarray}
dE/dx &\to& \frac {\sqrt {GM}}{ln_e {10}}dW/dx \\
E     &\to& \frac {\sqrt {GM}}{ln_e {10}}W.
\end {eqnarray}
Apart from the factor $\sqrt {GM}/ln_e {10}$, which is a normalisation factor,
$dE/dx$ may be interpreted as the modulus of a Breit-Wigner resonance with
$x$-dependent width:
%Eq. (44)
\begin {equation}
BW = \frac {\Gamma (x)/2}{E - E_F - i\Gamma (x)/2}.
\end {equation}   
There is a quantum mechanical pole at $E = E_F$ where $dW/dx$ varies fastest.
Astrophysical situations would be too noisy to reveal the phase variation.
However, it is questionable what to take as the absolute zero of
energy because of possible renormalisation effects, like those
responsible for the Lamb shift.  
The energy $W$ starts at zero because of local gauge invariance, and its
central value is shifted downward by 0.25 GM.
This accounts for the form of Eq. (8) of Section 2, where two terms
$(a_0/2a)^2$ cancel.

Suppose the graviton acquires an effective mass $\xi (x)$.
The Yukawa potential for gravitational coupling between two particles
of mass $m_1$ and $m_2$ is $V(r) = -Gm_1m_2 e^{-\xi (x)r}/r$, where the
exponential accounts for the graviton mass. 
From the observed variation of $dW/dx$, the exponential falls to 
half-height for a displacement $\Delta x = \pm 0.48$ on Fig. 1(c).

In subsection 2.1, the dependence of the fit on alternative forms of
Milgrom's $\mu$ function was tested.
Although acceleration curves change significantly, the Fermi function
remains rather stable. 
It changes significantly only at the ends of the range $x=-8$ to $-12$.
In Section 3, the stability of the fit using the empirical form of
$W$ in Eq. (24) has also been studied.
The conclusion is again that the shape of the Fermi function in Fig. 1(c) 
is very stable against these changes.

The conclusion from these results is that the central part of the Fermi
function is stable, but can be perturbed at the edges.
In superconductors, a coherence length was introduced by Pippard to
account for the effects of defects beyond an experimentally observed
range \cite {Pippard}. 
It appears that galaxies behave similarly.

An attempt was made by t'Hooft and Veltman to construct a renormalisation
scheme for gravitation, but was inconclusive \cite {t'Hooft}; 
that problem persists today.
It is obviously of interest to re-examine this problem in the extreme
infra-red where gravitons interact with nucleons to form galaxies.
One of the points for investigation is that an effective mass for
the graviton introduces a separation between the $s$, $t$ and $u$ channels
of the Mandelstam diagram, making it resemble Chiral Symmetry Breaking;
for massless gravitons, the three channels meet at a point.
This problem needs to be studied using the Schwinger-Dyson approach
along the lines of Wilson and Pennington \cite {Pennington}, but is a
non-trivial problem. 
It requires a full understanding of how the nucleon acquires its mass
from coupling to 3 or more quarks.
Presently, there are no precise Lattice Gauge calculations of the
nucleon mass.
There is however an alternative approach. 
Thomas developed a model in 1983 explaining features of deep inelastic
scattering (i.e. large momentum and energy transfers in electron
scattering from nucleons) which provided a qualitative explanation
of the data in terms of a chiral quark model \cite {Thomas}.
This has been followed up in a recent paper of Burkardt et al. which 
improves this work in specific ways \cite {Burkardt}.
Quoting the abstract of this paper: `The results pave the way for
phenomenological applications of pion cloud models that are manifestly
consistent with the chiral symmetry properties of QCD'.

Suppose the graviton does acquire an effective mass.
A feature of $\pi N$ elastic scattering is that there is a nucleon pole 
term at $s = M^2_n$ and an Adler zero at $s = u$, $t=m^2_\pi$.
It would be interesting to consider the effect of these perturbations
on the behaviour of the nucleon using the approach of Burkardt et al.
At first sight it appears that the form factor in the gravitational
potential will weaken gravity.
However, what one must remember is that the wave-lengths of gravitons
on a galactic scale can be very large, easily larger than a star.
The graviton then behaves nearly as a plane wave and is a coherent 
amplitude over a large range.
A detail is that a massive graviton does develop a scalar component.
A coherent amplitude can function like Cooper pairs in a superconductor and 
produce an energy gap.
A coherence length like that introduced by Pippard can arise in many ways.
Supernovae act as major perturbations, heating sizable volumes.
It is also known that so-called chimneys and wormholes provide
channels through which currents of dust and gas flow.
Furthermore, one of the remarkable features of galaxies is that they only 
grow to a certain size.
The largest have masses of order $10^{14} M_{\odot}$.
Clusters of galaxies do not assemble into one super-galaxy.
All of these effects point to a coherence length associated with
the variation of the Fermi function in Fig. 1(c).

These remarks are frankly speculative, but deserve further study.
There have been papers by van Dam \cite {Dam} and 
Boulware and Deser \cite {Deser} which claim no-go theorems preventing 
the graviton developing a mass, but it is not clear whether the present
situation evades these theorems.

Let us return to a simpler issue, the missing lower branch of the
Breit-Rabi equation. 
On this branch both $W$ and $dW/dx'$ change sign.
This change of sign requires that this branch describes an excited
state rather than a condensate.
Such an excited state is likely to decay on a time scale much less than
that of galaxies, so it is unlikely that this branch will be observable.
For those wishing to investigate this branch, 
the procedure is (a) to fit the upper branch as a function of $x$,
(b) rotate to $x',y'$ axes, (c) reverse the sign of $y'$ to reach the
lower branch, and (d) rotate axes back again to $x,y$.
The best place to search for this branch is near the crossing point
of Fig. 1(a).

Presently, Cold Dark Matter (CDM) is used by cosmologists as the source of
condensation.
However, it is not clear whether CDM is cored or cusped. 
Its radius beyond the peripheries of galaxies is not well established.
It might also be warm. 
Recently, Boylan-Kolchin, Bullock and Kaplinghat have made a study of
the condensation of dwarf galaxies using the standard $\Lambda CDM$ model
\cite {Boylan}.
They find that this model performs badly, even when mechanisms
such as supernova feedback are included.
Weinmann et al. reach very similar conclusions \cite {Weinmann}. 
An obvious question is what the Dark Matter is that has survived
for the age of the Universe without decaying.
If it did decay significantly, the evolution of galaxies would
change as a function of their age.
Mond and the scheme presented here do not have this problem. 

\subsection {The relation to Dark Energy}
If MOND successfully models the formation of galaxies and globular
clusters, it raises the question of how to interpret Dark Energy.
In a de Sitter universe, the Friedmann-Robertson-Walker model  
smoothes out structures using a $\Lambda CDM$ function which models the 
gross features.
These change over the lifetime of the Universe.
In this approach, the Unruh temperature \cite {Unruh} is very low, 
$5 \times 10^{-31} K$, because it is smoothed over the time taken for the 
Hubble acceleration to reach the velocity of light.
However, if quantum mechanics governs individual galaxies, there will
instead be fine structure in Dark Energy. 
It is logical that steps like Figs. 1(c) do not just average out, but 
instead accumulate over all of this fine structure.
The sum total of the fine structure is what contributes to the 
conventional Friedmann-Robertson-Walker model of the Universe.
This is parametrised via the assumed time dependence of the metric on the 
Hubble acceleration.
A similar suggestion along these lines has been advanced by Zhang and Li
\cite {Zhang} using ideas based on entropic arguments. 
The `present-day' Hubble acceleration is the local value and varies over
the Universe according to the parametrisation by Dark Energy.

There is a further argument pointing towards the idea that local fine
structure is cumulative.
Peebles and Nusser argue that galaxies condense more rapidly than the 
standard $\Lambda CDM$ model predicts \cite {Nusser}.
In particular, they point out that the Local Void contains far
fewer galaxies than $\Lambda CDM$ predicts statistically,
while there is an unexpected presence of large galaxies on the
outskirts of the Local Void.
Their Fig. 1 is very persuasive in this respect.
Only 3 galaxies are observed in the Local Void compared with 19
predicted.  
The Poisson probability for this result 
is of the order $10^{-5}$ from $\Lambda CDM$.
Peebles and Nusser conclude: `In short, the general sensitivity of
galaxies to their environment is not expected in standard ideas.
It would help if galaxies were more rapidly assembled so that they
could then evolve as more nearly isolated island universes.'
Later Peebles considered an additional empirical term added to the
$\Lambda$ QCD model, but comments that the change requires that
Cold Dark Matter is cored rather than the expected cusped behaviour
\cite {Peebles}.

A natural explanation is that the Local Void gives no contribution
to the Hubble mechanism, except for its three galaxies.
The total energy $E$ is then higher there.
On the periphery of the Void, there is gas and dust which can form
galaxies.
This gas runs down the energy gap to enlarge galaxies forming there.
It is likely that in the early Universe dwarf galaxies form first;
globular clusters are known to be very old and may be relics of
such small galaxies. 
Galaxies then grow by accumulation of dust and gas.
Large galaxies are known to spin off small ones in collisions.  

\section {Further work which is needed}
The model proposed here is precise and open to experimental test. 
The most important and simplest is that if Dark Matter is replaced by 
this model, it is obviously necessary to redo the parametrisation of Dark 
Energy so that it reproduces smoothly what has been parametrised as Dark 
Matter up to now.
This does not necessarily require major modifications.
The main point is to fit the third peak in the spectrum.
This is an exercise requiring cooperation of groups with the 
latest data and techniques at their finger-tips.
Despite the clues pointed out here that a condensate explains
galactic rotation curves, this could fail badly.
It would be hardly surprising if minor modifications are required
in the refit to Dark Energy.

Angus, Famoey and Buote have considered the possibility that neutrinos may 
condense within the gravitational potentials of galaxies \cite {Angus}.
This is possible, but it is difficult to be quantitative at present because
only relative masses of neutrinos are known, not absolute values;
also there is a speculative possibility that sterile neutrinos exist, but 
that is still an open question.

The mixing amplitude  will affect both Strong and Weak
Gravitational Lensing. 
However, unless data are used where a distant quasar transmits light 
through the periphery of a galaxy, where the effect of $W(x)$ is large, 
the lensing effect will be rather small.
The weak logarithmic tail of Newtonian gravitation also needs to be taken 
into account.

What happens in clusters of galaxies needs detailed, laborious
calculations using the formulae given here; since there are presently
claims that Milgrom's formula does not correctly reproduce what happens
in clusters, this should be revealing.
A multi-body interaction between the Hubble mechanism and several 
galaxies is required, i.e. a generalisation of the Breit-Rabi equation.
Quite apart from the attraction to acceleration $a_0$, 
there are also strong tidal effects of the variety discussed by
Angus {\it et al.} \cite {Angus} and Kroupa \cite {Kroupa}.
 
The same remarks apply to the Bullet Cluster, which needs to be refitted 
with the model proposed here.
The calculation becomes a two-centre problem. 
The Hubble acceleration (at red-shift $z \simeq 0.2$) couples to
both galaxies, but they also couple to one another, modifying
the zero-point energy;  
individual stars in galaxies communicate with one another not only
through the Newtonian potential but via their Fermi functions.
In the Bullet Cluster, each galaxy behaves in this way, but there will
be coupling between stars "belonging" to each  individual galaxy.
There may be complex interactions between the two galaxies including 
resonance effects. 
There are new data on the Bullet Cluster \cite {Paraficz}  which may 
offer new interpretations. 

\section {Concluding Remarks}
The equations given here allow very little freedom - just the $4\%$ 
perturbations allowed in the Fermi function at the top and bottom of
Fig. 1(c). 
There are five clues which point to galaxies and globular clusters being
quantum mechanical condensates.
Firstly, the phenomena of Fig. 1 appear on a log-log plot, as expected
from Statistical Mechanics;
Fig. 1(a) can be interpreted in terms of quantum mechanical mixing
between two crossing eigenstates.
Secondly, Fig. 1(c) is fitted naturally by a Fermi function with 
an energy gap $0.5\, GM$.
Thirdly, the asymptotic form of the acceleration in Fig. 1(a) generates
a logarithmic tail; this requires a quantum mechanical explanation.
Fourth, the rotation of axes accomodated by Fig. 3 is just what is
expected from the Boguliubov transformation for a condensate.
A Fermi-Dirac condensate fits the data; a Bose-Einstein condensate 
does not.
Fifth, the interpretation of the curve of Fig. 1(b) as an
energy-dependent Breit-Wigner pole
suggests a connection with Chiral Symmetry breaking.
It suggests that the graviton acquires an effective 
mass in the vicinity of $a_0$.
The precise mechanism is a challenge to theorists and is presently a
conjecture. 
A final point is that the lower branch of the Breit-Rabi equation has
the opposite curvature to the upper one; the natural interpretation of
this result is that it corresponds to an excited state which will decay
rapidly and will not therefore appear in galactic rotation curves.
An unanswered question is why the asymptotic form of the acceleration
is $\sqrt {a_0g}$.
This is taken from experiment. 
 
The standard approach to the Universe is the Friedmann-Robertson-Walker 
model where the metric is smoothed over local structures and appears in 
the metric of General Relativity.
My suggestion is that galaxies create fine-structure and
the Friedmann-Robertson-Walker model becomes the sum 
over these structures.
This would explain the agreement between $a_0$ and $cH_0/2\pi$ in our
part of the Universe.

Those are the essential points. They do reflect successfully the facts 
which MOND parametrises and they do fit together in a way familiar in
Particle Physics in a condensate due to Chiral Symmetry Breaking.  

There are two further small comments.
Firstly, there have been many papers on the derivation of Newton's law from
Entropic arguments. 
Many assume that one can assign a temperature $T_0$ to the Hubble
acceleration and a temperature $T_1$ for the gravitational potential 
by adding them as the sum of squares, i.e. the random phase 
approximation. 
This gives the result for the total acceleration:
%Eq. (45)  
\begin {equation}
a = \sqrt {(g_N + g_{H})^2 - g_H^2}
\end {equation}
where $g_H$ is the Hubble acceleration $H_0^2 r$, and $r$ is the
distance to the centre of the gravitating object. 
This is a quite different approach to the one proposed here and
leads to a very small effect. 
For the Milky Way, it is a maximum of $2 \times 10^{-4} cH_0/2\pi$.

Secondly, Iorio has calculated that effects of MOND on the perihelia of 
planets in the Solar system are about a factor 10 below present 
experimental errors \cite {Iorio}. 
The logarithmic term arising from MOND will be of order $a_0$, which
is very small and will vary exceedingly slowly over the solar system.
It has been pointed out by Galianni et al. that it may be feasible to
detect the MOND effect at acceleration $a_0$ in the solar system near 
Lagrangian points where the accelerations of the Sun, Earth and Moon 
cancel out \cite {Galianni}.
The LISA Pathfinder space mission A is planned to visit this area in 
2013.
If measurements of sufficient accuracy could be made,
they might confirm or modify MOND as a model of the behaviour of galaxies; 
secondly, they have the important potential to measure the shape of the 
response function through the region where the bend appears in Fig. 1. 
The alternative approach is the study of globular clusters, which 
provide a different perspective to the rotation curves of galaxies.
\vskip 3mm
\noindent {\small ACKNOWLEGEMENT}
\newline
I wish to thank Prof. Pedro Bicudo for discussions about Bose-Einstein 
condensates.

\end {document}